\documentclass[10pt]{iopart}

\usepackage{color}
\usepackage{iopams}
\usepackage{bm}
\usepackage{multirow}
\usepackage{dsfont}
\usepackage{graphicx}
\usepackage{enumerate}
\usepackage[square,sort&compress,numbers]{natbib}
\usepackage{units}
\usepackage{mathptmx, textcomp}
\usepackage{epstopdf}
\usepackage{amssymb}
\usepackage{etoolbox}
\usepackage{hyperref}

\makeatletter
\def\@mkboth#1#2{}
\newlength\appendixwidth
\preto\appendix{\addtocontents{toc}{\protect\patchl@section}}
\newcommand{\patchl@section}{%
  \settowidth{\appendixwidth}{\textbf{Appendix }}%
  \addtolength{\appendixwidth}{1.5em}%
  \patchcmd{\l@section}{1.5em}{\appendixwidth}{}{\ddt}%
}
\makeatother

\newcommand{\mri}{\mathrm{i}}
\newcommand{\Exp}[1]{\mathrm{e}^{\mbox{\footnotesize$#1$}}}
\newcommand{\ket}[1]{|#1\rangle}

\renewcommand{\vec}[1]{\mathbf{#1}}

\def\ket#1{\left|#1\right\rangle}
\def\39K{$^{39}$K}
\def\87Rb{$^{87}$Rb}

\begin{document}
\title{Time-of-flight expansion of binary Bose-Einstein condensates at finite temperature}

\author{K. L. Lee$^1$, N. B. J\o rgensen$^2$, L. J. Wacker$^{2, 3}$, M. G. Skou$^2$, K. T. Skalmstang$^2$, J. J. Arlt$^2$ and N. P. Proukakis$^1$}

\address{$^1$ Joint Quantum Centre (JQC) Durham-Newcastle, School of Mathematics, Statistics and Physics, Newcastle University, Newcastle upon Tyne NE1 7RU, England, UK}
\address{$^2$ Institut for Fysik og Astronomi, Aarhus Universitet, Ny Munkegade 120, 8000 Aarhus C, DK}
\address{$^3$ Danish Fundamental Metrology, Kogle All\'{e} 5, 2970 H\o rsholm, DK}

\eads{\mailto{nikolaos.proukakis@newcastle.ac.uk}}
\date{\today}

\begin{abstract}\noindent
Ultracold quantum gases provide a unique setting for studying and understanding the properties of interacting quantum systems. Here, we investigate a multi-component system of $^{87}$Rb--$^{39}$K Bose-Einstein condensates (BECs) with tunable interactions both theoretically and experimentally. Such multi-component systems can be characterized by their miscibility, where miscible components lead to a mixed ground state and immiscible components form a phase-separated state. Here we perform the first full simulation of the dynamical expansion of this system including both BECs and thermal clouds, which allows for a detailed comparison with experimental results. In particular we show that striking features emerge in time-of-flight for BECs with strong interspecies repulsion, even for systems which were separated {\em in situ} by a large gravitational sag. An analysis of the centre of mass positions of the BECs after expansion yields qualitative agreement with the homogeneous criterion for phase-separation, but reveals no clear transition point between the mixed and the separated phases. Instead one can identify a transition region, for which the presence of a gravitational sag is found to be advantageous. Moreover we analyse the situation where only one component is condensed and show that the density distribution of the thermal component also show some distinct features. Our work sheds new light on the analysis of multi-component systems after time-of-flight and will guide future experiments on the detection of miscibility in these systems.
\end{abstract}



\maketitle

%*******************************************************************************************************
\section{Introduction}
%*******************************************************************************************************

The non-equilibrium dynamics of interacting quantum systems is a non-trivial and highly relevant field of research, since it has direct impact on our ability to control such systems in the next generation of quantum devices. Ultracold quantum gases are attractive systems to study such phenomena~\cite{weiner_bagnato_1999, dalfovo_giorgini_1999, bloch_dalibard_2008}, since they provide high purity, in tailored potentials~\cite{grimm_weidemuller_2000, gaunt_schmidutz_2013}, with controllable interactions~\cite{chin_grimm_2010}. Moreover they allow for the realization of multi-component homo- and heteronuclear systems~\cite{myatt_burt_1997, hall_matthews_1998, maddaloni_modugno_2000, papp_pino_2008, sugawa_yamazaki_2011, modugno_modugno_2002, thalhammer_barontini_2008, mccarron_cho_2011, lercher_takekoshi_2011, pasquiou_bayerle_2013, wacker_jorgensen_2015, wang_li_2016, ferrier-barbut_delehaye_2014}, spinor systems~\cite{stamper-kurn_ueda_2013}, and dipolar mixtures~\cite{Maier2015,Ilzhofer2017}.

A key property of multi-component systems, is their miscibility~\cite{esry_greene_97, pu_bigelow_1998, ao_chui_98,timmermans_98, ohberg_1999, trippenbach_goral_2000, delannoy_murdoch_2001, riboli_modugno_02, jezek_capuzzi_02, kasamatsu_tsubota_2004, ronen_bohn_2008, takeuchi_ishino_2010, suzuki_takeuchi_2010, mason_aftalion_2011, aftalion_mason_2012, wen_liu_12, pattinson_billam_2013, hofmann_natu_2014, pattinson_parker_2014, liu_pattinson_2016, white_hennessy_16}. Miscible components overlap in space leading to a mixed ground state, while immiscible components form a phase-separated state. Experimentally, it is often difficult to detect the miscible-immiscible transition by direct {\em in situ} imaging of two trapped components. Therefore most experiments have investigated the transition between mixed and separated phases by observing the time-of-flight (TOF) expansion of a mixture of Bose-Einstein condensates (BECs)~\cite{hall_matthews_1998,papp_pino_2008,wacker_jorgensen_2015,wang_li_2016,mccarron_cho_2011}. These experiments have typically identified the transition by observing a dramatic increase in the separation between the centre of mass (COM) positions as the interaction strength was changed, with associated strong features at the interface between the expanded mixtures. The observed transition points were found to be in good quantitative agreement with those predicted by the criterion for the transition from miscible to immiscible in a homogeneous system. This criterion is given by the miscibility parameter $\Delta=(g_{11}g_{22}/g_{12}^2) - 1$ which depends on the intra- ($g_{11}$, $g_{22}$) and interspecies ($g_{12}$) interaction strengths and determines whether two components mix ($\Delta>0$) or phase-separate ($\Delta<0$). 

However, we have previously shown that this criterion is not applicable in the trapped case and proposed an alternative criterion based on the {\em in situ} density distributions, which can be probed through the observation of dipole oscillations~\cite{lee_jorgensen_2016}. These simulations further indicated that the interactions significantly affect the expansion dynamics of these binary systems, and hence details of  the relation between the \emph{in-situ} density distributions and the COM positions after TOF expansion remained unclear.

In this work, we perform the first full simulation of the dynamical expansion of a  multi-component system including both BECs and thermal clouds and compare these results with experiments on a mixture of $^{87}$Rb--$^{39}$K atoms. We show that striking features emerge in TOF for BECs with strong interspecies repulsion, even for two systems which were geometrically separated {\em in situ} by their large trap offsets. Moreover we show that  a measurement of the centre of mass position of binary BECs after TOF expansion yields a result in qualitative agreement with the homogeneous criterion for phase-separation. In both cases we obtain good agreement between our simulation and the experimental results. This reconciles our previous theoretical analysis~\cite{lee_jorgensen_2016}, with experimental observations~\cite{hall_matthews_1998,papp_pino_2008,wacker_jorgensen_2015,wang_li_2016,mccarron_cho_2011}.

In addition our analysis reveals a multitude of other interesting features. If only one component is condensed, the density distribution is shown to retain distinct features. Although the BEC in the partly-condensed component is found to be largely symmetric after expansion, the non-condensed species exhibits asymmetric features, which are enhanced in the immiscible case. Moreover, in the case of both components being partly-condensed, we show that a trap offset (e.g. due to a gravitational sag) is useful in yielding a pronounced shift of the COM position after expansion, thus facilitating an easier characterization of the properties of mixtures of different species. This was in fact implicitly used already in early binary mixture experiments (see e.g. Fig.~2 in Ref.~\cite{hall_matthews_1998}). Finally, we show that there is no abrupt transition which can be observed in TOF. 

In the remaining sections, we first outline the experimental procedure used to investigate binary mixtures (Sec.~\ref{sec:exp_imp}) and give an overview of the theoretical tools used to model this scenario (Sec.~\ref{sec:theory}). We then perform a detailed analysis of the spatial distributions and COM positions of binary mixtures after TOF expansion  (Sec.~\ref{sec:tof}). This includes the effect of temperature and of the gravitational sag on TOF expansion. We summarize  our key findings in section~\ref{sec:conclusion}.

%*******************************************************************************************************
\section{Experimental and Theoretical Methodology}
%*******************************************************************************************************

In the following a detailed description of the experimental production and detection of binary BECs is given. Moreover we provide an overview of the theoretical methods which enable us to perform a full numerical simulation of the TOF expansion of binary BECs in the presence of a thermal clouds. Further technical details are given in~\ref{sec:app_numerical}.

%*******************************************************************************************************
\subsection{Experimental realization of binary BECs}
\label{sec:exp_imp}
%*******************************************************************************************************

Our experiments with binary mixtures are performed with $^{87}$Rb and $^{39}$K atoms in an optical dipole trap. The optical dipole trap allows for free tuning of the external magnetic field and thus Feshbach resonances can be employed to tune the scattering length between the two species. Moreover the optical trap allows for particularly rapid switching of the external potential and thus a precise comparison of the TOF dynamics with theoretical results is possible. However, due to the different masses, the gravitational sags of the two species differ significantly. The effect from this relative trap offset is analysed further in Sec.~\ref{sec:sag}.

\begin{figure}[htb]
\begin{center}
\includegraphics[width=0.8\textwidth]{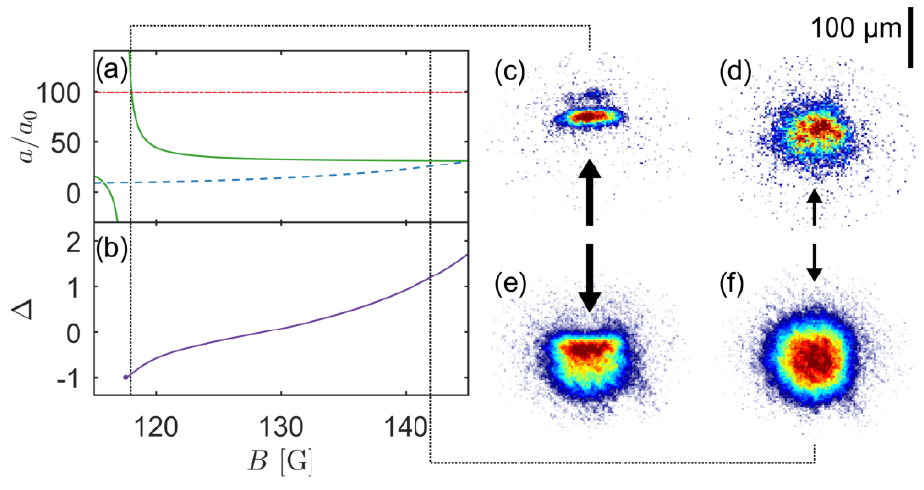}
\caption{Scattering properties of \39K and \87Rb atoms, and the homogeneous miscibility parameter $\Delta=g_{11}g_{22}/g_{12}^2-1$. (a) Scattering lengths of \39K (dashed blue), \87Rb (dashed-dotted red) and between the two species (solid green). (b) The miscibility parameter ranges from miscible $\Delta>0$ to immiscible $\Delta<0$. The two dotted lines indicate immiscible $\Delta\approx -1$ (left) and miscible $\Delta\approx 1$ (right) parameters. (c-f) Examples of measurements at these parameters, where (c,d) shows \39K and (e,f) \87Rb. In the immiscible case (c,e) the density distributions and the centre of mass positions of the two components are strongly modified.}
\label{fig:introFig}
\end{center}
\end{figure}

Figure~\ref{fig:introFig} shows the relevant scattering properties. We perform the experiments with both species in the $\ket{F=1,m_F=-1}$ state in a range of magnetic fields where the scattering length of \39K is positive, allowing for the production of binary BECs~\cite{wacker_jorgensen_2015}. Moreover an interspecies Feshbach resonance is available in this region and thus the miscibility parameter can be tuned across the transition point. At a given magnetic field $B$, the s-wave scattering length between \39K atoms is described by
\begin{equation}
a_{\rm K-K} (B) = -19a_0\left(1-\frac{\unit[55]{G}}{B-\unit[33.6]{G}} + \frac{\unit[37]{G}}{B-\unit[162.35]{G}}\right),
\label{eqn:aKK}
\end{equation}
and the scattering length between \39K and \87Rb atoms is
\begin{equation}
a_{\rm Rb-K} (B) = 29.7a_0\left(1+\frac{\unit[1.21]{G}}{B-\unit[117.56]{G}}\right),
\label{eqn:aKRb}
\end{equation}
where $a_0$ is the Bohr radius~\cite{wacker_jorgensen_2015, roy_landini_2013}. The scattering length between \87Rb atoms is constant at $a_{\rm Rb-Rb} = 99a_0$. The interaction strengths are related to the scattering lengths as $g_{ij} = 2\pi\hbar^2 a_{ij} (m_i + m_j )/(m_im_j)$, where the indices refer to the two species with respective masses $m_i$. This allows us to study mixtures with miscibility parameters $\Delta=g_{11}g_{22}/g_{12}^2-1$, from highly-miscible ($\Delta \sim 1$) to highly-immiscible ($\Delta\sim-1$), as shown in Fig.~\ref{fig:introFig}.

The details of the experimental apparatus and procedure are described in~\cite{wacker_jorgensen_2015} and briefly summarized in the following. Initially, a dual-species magneto-optical trap simultaneously captures both species of atoms from a background vapour. This is followed by an optical molasses which cools them to sub-Doppler temperatures. Afterwards both species are optically pumped into the fully stretched $\ket{2,2}$ state. The atoms are loaded into a magnetic quadrupole trap which transports them to a separate chamber where the remainder of the experimental procedure is performed. Here, they are loaded into a different quadrupole trap and evaporative cooling is performed. Microwave radiation is used to selectively cool \87Rb atoms, which cool \39K sympathetically. Next, an additional coil transforms the trap into a harmonic Ioffe-Pritchard configuration where further microwave cooling is performed.

Before the atoms reach quantum degeneracy, they are loaded into an optical dipole trap consisting of two crossed laser beams at a wavelength of $\unit[1064]{nm}$. A rapid adiabatic passage using radio frequency radiation transfers both species from the $\ket{2,2}$ to the $\ket{2,-2}$ state. From here a radio frequency $\pi$-pulse and a rapid adiabatic passage transfer \87Rb and \39K atoms into the target state $\ket{1,-1}$ respectively.

The final evaporation is performed by lowering the power of the dipole trap laser, which preferably allows \87Rb to leave the optical trap due to its larger gravitational sag. The rethermalisation of the two species during evaporation, is enhanced by setting the magnetic field to $\unit[118.7]{G}$, which is in the vicinity of the interspecies Feshbach resonance. After evaporation, the field is adjusted to a desired target value, and the dipole trap power is increased to ensure further rethermalisation, resulting in trap frequencies of $(\omega_\rho, \omega_z)_{\rm K} /2\pi = \unit[(132, 186)]{Hz}$ and $(\omega_\rho, \omega_z)_{\rm Rb} /2\pi = \unit[(93, 124)]{Hz}$ for \39K and \87Rb, respectively. At these trap parameters the gravitational sag separates the centres of the potentials for the two species leading to an estimated trap offset of $7.1\pm\unit[0.7]{}$\textmu$ \rm{m} $.

Finally, the TOF detection is realized by turning off the optical trap, and allowing the atoms to expand freely. The magnetic field is kept at the desired value for $\unit[8]{ms}$ allowing the atoms to interact during the initial expansion where interactions are significant. Subsequently the field is turned off and absorption images of \39K and \87Rb atoms are taken after $\unit[23]{ms}$ and $\unit[24.7]{ms}$  respectively. The time between these images is set by technical limitations of the camera system. Examples of absorption images of the two components are shown in Fig.~\ref{fig:introFig}~(c-f).

By tailoring the evaporation sequence in the dipole trap appropriately, it is possible to condense both species, only \87Rb, or none of the two. When condensing both species, we typically obtain about $3\times10^4$ \39K and $1.2\times10^5$ \87Rb atoms at a temperature of $150$--$\unit[200]{nK}$, resulting in BECs of respectively $8\times10^3$ \39K and $5\times10^4$ \87Rb atoms. The sum of the Thomas-Fermi radii of the condensates in the vertical direction is approximately 6\,\textmu$ \rm{m}$, which is on the order of the estimated trap offset.

To evaluate how the interspecies interaction affects the density distributions, we additionally record images of each species without the other one present. This allows us to evaluate the undisturbed COM position and spatial distribution of each species. These undisturbed COM positions are used as the origin of the coordinate systems for each species in all experimental data shown.

%*******************************************************************************************************
\subsection{Theoretical model}
\label{sec:theory}
%*******************************************************************************************************

The finite temperature dynamics of a partially-condensed system is well described in the context of a ``two-gas'' model (``Zaremba-Nikuni-Griffin'', or ``ZNG'' model \cite{zaremba_nikuni_1999,Griffin2009}), consisting of a BEC and a thermal cloud. Here we follow our earlier work~\cite{edmonds_lee_2015a, edmonds_lee_2015b, lee_proukakis_16} which appropriately generalized this to a binary mixture in such a way that both BECs and both thermal clouds are coupled within and between the mixture components through both mean-field and collisional interactions. In order to probe expansion dynamics, we limit our discussion here to the collisionless limit of the above theory: in this limit, each BEC is described by a generalized Gross-Pitaevskii equation (GPE),
\begin{equation}
i\hbar\frac{\partial\phi_j}{\partial t}=\bigg[-\frac{\hbar^2}{2m_j}\nabla^2+U^{j}_{c}\bigg]\phi_j,\label{eq:gpe}
\end{equation}
where $j$ denote the component. The thermal clouds are described by a \emph{collisionless} quantum Boltzmann equation
\begin{equation}\label{eq:qbe}
\frac{\partial}{\partial t}f^{j}+\frac{1}{m_j}{\bf p}\cdot\nabla_{\bf r}f^{j}-\nabla_{\bf p}f^{j}\cdot\nabla_{\bf r}U^{j}_{\rm n}=0,
\end{equation}
where $f^j(\vec{p},\vec{r},t) = \int d\vec{r}' \Exp{\mri \vec{p}\cdot\vec{r}'/\hbar}\langle \delta^\dag\left(\vec{r}+\vec{r}'/2,t\right)\delta\left(\vec{r}-\vec{r}'/2,t\right)\rangle$
denotes the Wigner distribution of the thermal atoms 
(with $\hat{\delta}_j = \hat{\Psi}_j - \phi_j$ the fluctuation operator after the BEC mean field, $\phi_j$ has been subtracted).
The local thermal cloud density is then given by $\tilde{n}_j(\vec{r}) = \int d\vec{p}/(2\pi\hbar)^3 f^j(\vec{p},\vec{r},t)$ while the density of the whole cloud $n_{{\rm tot},j}$ is obtained as the sum of the two, i.e. $n_{{\rm tot},j}= n_{c,j} + \tilde{n}$, where $n_{c,j}=|\phi_j|^2$. The total number of atoms (which is fixed in our model) is obtained as $N_j = \int d\vec{r}\;n_{{\rm tot},j}(\vec{r})$.

In contrast to the usual zero-temperature case, the BEC atoms now experience an effective potential $U_c^j$ corrected by the presence of the thermal cloud, while the thermal atoms are modelled as classical particles moving in an effective potential $U_n^j$. These potentials, which are found to play a key role during the early stages of the expansion process, include both the external potential $V_j$ and the mean-field contributions, which are related to the BEC density $n_{c,j}$ and the thermal density $\tilde{n}_j$, by
\begin{numparts}
	\begin{eqnarray}
		\label{eq:effUc}U_c^j =& V_j+g_{jj}(n_{c,j} +2\tilde{n}_j)+g_{kj}(n_{c,k} +\tilde{n}_k),\\
		\label{eq:effUt}U_n^j =& V_j+ 2g_{jj}(n_{c,j} + \tilde{n}_j) +g_{kj}(n_{c,k} + \tilde{n}_k).
	\end{eqnarray}
\end{numparts}

The trapping potentials used here are defined by
\begin{equation}
V_j(\vec{r})=\frac{1}{2}m_j\bigl[\omega_{\rho,j}^2 \rho^2 + \omega_{z,j}^2 (z-z_{{\rm s},j})^2\bigr]
\end{equation}
with trap centres $z_{{\rm s},j}$, and radial and axial angular frequencies, $\omega_{\rho,j}$ and $\omega_{z,j}$, respectively. The above expressions are consistent with standard multi-component Hartree-Fock theory~\cite{ohberg_stenholm_1998}, which includes an extra factor of 2 in the condensate mean-field potential (Eq.~(5a)) associated with the direct and exchange interactions involving thermal and condensate atoms of the same species. A corresponding factor of 2 is included in the total density contributions appearing in the potential for the thermal atoms (Eq.~(5b) within the same component. Such factors are however absent in the coupling {\em between} the components (final contributions in each of Eq.~(5a-b)).

The experimental scenario is simulated as follows. The initial equilibrium states are obtained in the usual way~\cite{jackson_zaremba_2002, edmonds_lee_2015a, edmonds_lee_2015b, lee_proukakis_16}. At time $t=0$, the atoms are released from the trap (i.e. $V_j(\vec{r})=0$) and expand freely. The expansion dynamics is simulated in a fixed two-dimensional grid utilizing the cylindrical symmetry of the problem which allows us to study atomic clouds that have expanded up to approximately ten times their in-trap sizes without changing the grid spacing. Further expansion is computationally prohibitive due to the need to accurately capture the length scales associated with the non-negligible expansion speeds, as discussed in \ref{sec:app_numerical}, which also gives further details on the implemented numerical approach. Similar to the experimental procedure where the magnetic field is changed during expansion for imaging purposes, the scattering length in the simulation is abruptly changed after $\unit[6]{ms}$, which is however found to have no noticeable effect on the COM position of the already diluted expanded clouds. Further details of our numerical implementation and the effect of scattering length change are given in \ref{sec:app_numerical} and \ref{sec:role_of_quench} respectively.

The simulation parameters are based on our previous experimental work~\cite{wacker_jorgensen_2015}. We thus consider a mixture of $1.2\times10^5$ $^{87}$Rb atoms and $4.2\times10^4$ $^{39}$K atoms in a harmonic trap with trapping frequencies of $(\omega_\rho, \omega_z)_{\rm K} /2\pi = \unit[(136, 189)]{Hz}$ and $(\omega_\rho, \omega_z)_{\rm Rb} /2\pi = \unit[(96.9, 129)]{Hz}$. In this situation the trap centres are separated by an offset of $6.9\pm\unit[0.5]{}$\textmu$ \rm{m}$ due to the gravitational sag. Both species are assumed to be at a relatively high temperature $T = \unit[200]{nK}$ (unless otherwise stated) for which there is approximately 15\% condensed fraction in $^{87}$Rb atoms and $26\textrm{--}30$\% condensed fraction in $^{39}$K atoms, with exact numbers slightly dependent on the precise values of interaction strengths. The scattering lengths are set by the magnetic field through Eqs.~(\ref{eqn:aKK})-(\ref{eqn:aKRb}), and both finite-size and mean-field corrections are taken into account when calculating the critical temperatures for the two species~\cite{dalfovo_giorgini_1999}.

In the following sections we compare our simulations to experimental data obtained according to section~\ref{sec:exp_imp}. The relevant parameters are similar, but the data was recorded using considerably longer expansion time, which allows spatial features to be resolved more precisely. Consequently, the experimental density distributions are approximately twice as large as the simulated distributions. However, this has no influence on main arguments and conclusions. Section~\ref{sec:COMshift} is based on a comparison with experiments performed at relatively short time of flight which were previously presented in~\cite{wacker_jorgensen_2015}. 

For displaying our theoretical results, we use the centre between the two traps as zero point, unless otherwise stated.

%*******************************************************************************************************
\section{Time-of-flight investigation of miscibility}
\label{sec:tof}
%*******************************************************************************************************

The experimental realization of binary BECs of $^{87}$Rb and $^{39}$K in an optical dipole trap and the availability to tune the interaction strength allow for a detailed experimental investigation of miscibility in this system. Our simulations, which include the full dynamical evolution of the two density distributions and their mutual mean-field effects, can thus be compared to the experiment directly.

%*******************************************************************************************************
\subsection{Density distributions of binary Bose-Einstein condensates after expansion}
\label{sec:distribution}
%*******************************************************************************************************

\begin{figure}[htb]
\begin{center}
\includegraphics[width=0.7\textwidth]{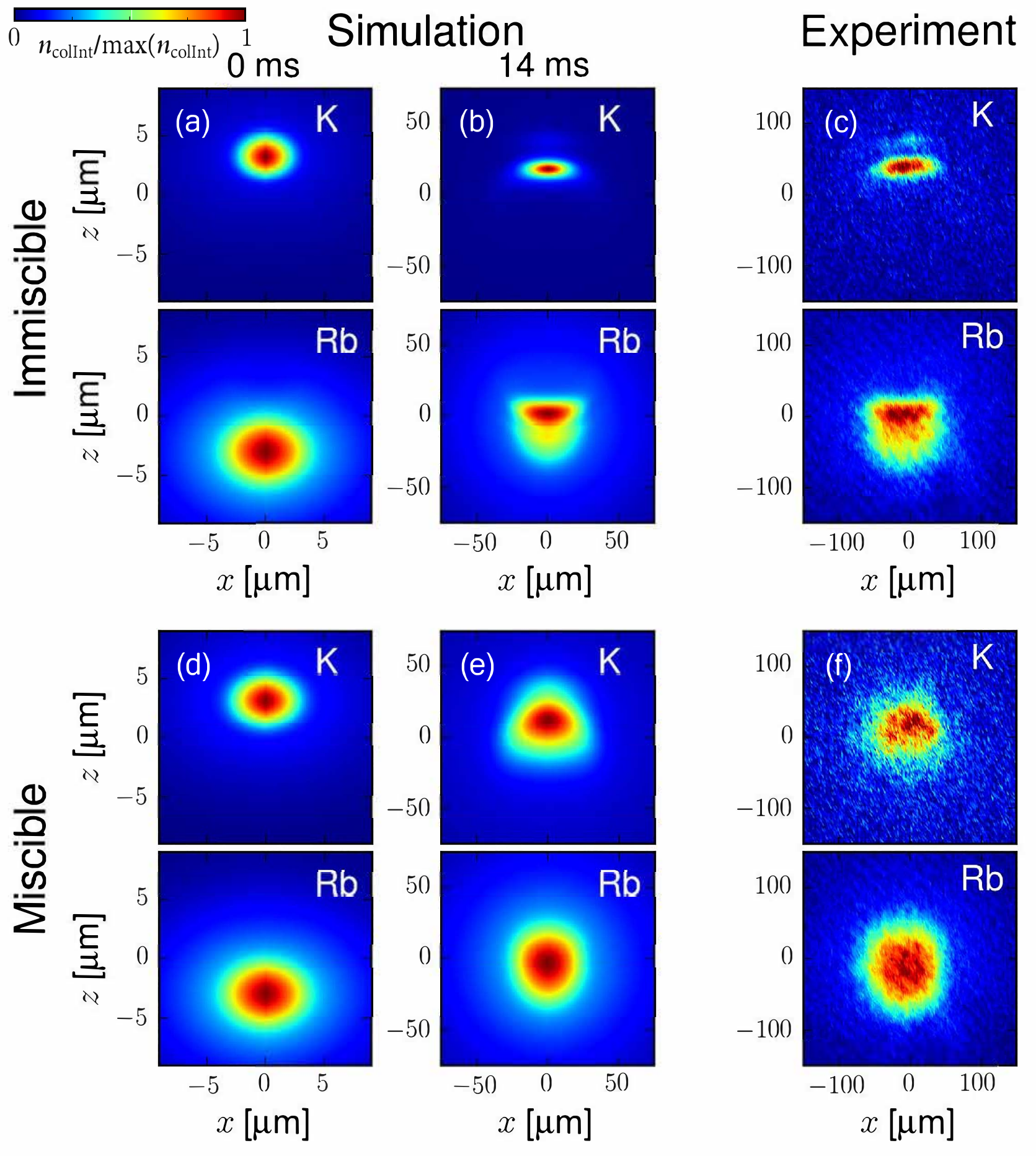}
\caption{Column-integrated density distributions $n_{\rm colInt}$ of $^{87}$Rb and $^{39}$K atoms. Simulations \emph{in-situ} (left column) and after 14\,ms of TOF expansion (middle column). Experimental TOF images after $\unit[23]{ms}$ (\39K) and $\unit[24.7]{ms}$ (\87Rb) (right column). Immiscible case ($\Delta = -0.93$ in simulations and $\Delta = -0.98$ in experiments) (top row) and miscible case ($\Delta = 1.2$ in simulations and experiments) (bottom row). The coordinate system of the simulated results corresponds to a freely falling frame initially centred between the two trap centres. The coordinate system of the experimental results for each species is centred on the position of the freely expanding BEC without the presence the other species.}
\label{fig:tof_b4_after}
\end{center}
\end{figure}

The simulated density distributions before and after TOF expansion for both a miscible and an immiscible mixture, are shown in Fig.~\ref{fig:tof_b4_after}, alongside the corresponding experimental images. A visual comparison of the distributions after TOF in the miscible and immiscible limits, immediately shows the good agreement between theory and experiment.

The \emph{in-situ} distributions do not show any distinguishing feature between the miscible and immiscible mixtures for the chosen trap. The atomic clouds are generally elliptical with widths on the order of the single-component Thomas-Fermi radii, which have a larger value along the $x$-axis due to of the smaller radial trap frequencies. This lack of features arises from the fact that the BECs are separated by a distance, which is similar to the sum of the Thomas-Fermi radii.

\begin{figure}[htb]
\begin{center}
\includegraphics[width=0.9\textwidth]{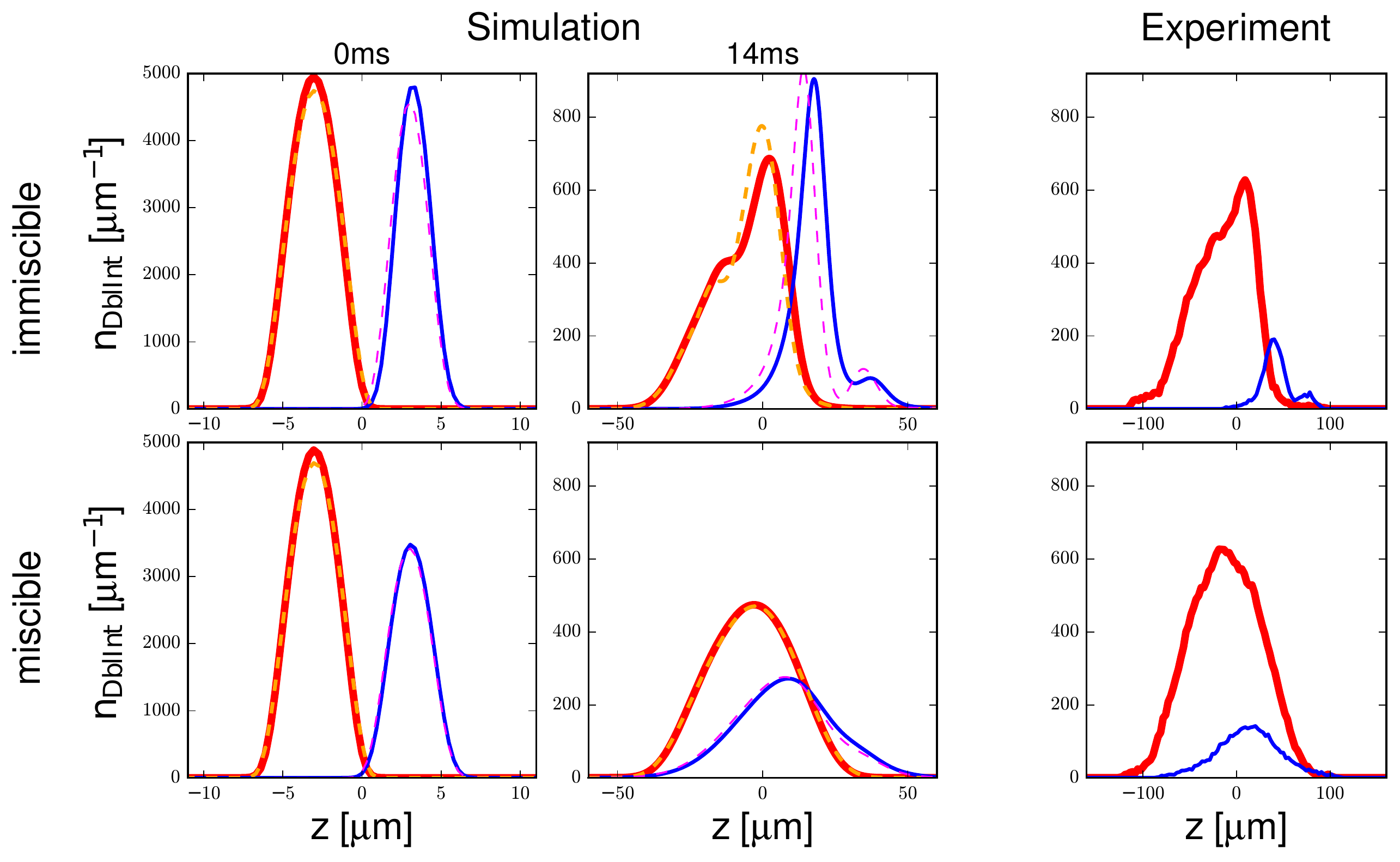}
\caption{Doubly-integrated density distributions $n_{\rm DblInt}$ of $^{87}$Rb (red solid line) and $^{39}$K (blue solid line) atoms. Simulations \emph{in-situ} (left column) and after 14\,ms of TOF expansion (middle column). In addition the result of a $T=0$ GPE simulation for $^{87}$Rb (orange dashed line) and $^{39}$K (purple dashed line) is shown. Experimental distributions after $\unit[23]{ms}$ (\39K) and $\unit[24.7]{ms}$ (\87Rb) TOF expansion (right column). Miscibility parameters and coordinates systems as in~Fig.~\ref{fig:tof_b4_after}. Note that the simulation was conducted with a larger $^{39}$K BEC fraction. Nonetheless the relevant features are the same in simulation and experiment.
}
\label{fig:dblint_tof_b4_after}
\end{center}
\end{figure}

During the expansion of individual BECs the aspect ratio typically inverts~\cite{castin_dum_1996,Kagan1996}, leading to a larger width along the axial direction. However, the repulsion between the two species strongly alters the dynamics and a flat interface between the two species is observed in our simulations and experiments as shown in Fig.~\ref{fig:tof_b4_after}~(b,c,e,f). Similar structures have previously been observed in other studies~\cite{hall_matthews_1998, papp_pino_2008, wacker_jorgensen_2015, wang_li_2016, mccarron_cho_2011}. In particular in the case of an immiscible mixture, the $^{39}$K BEC is strongly compressed and deviates dramatically from the predicted size due to self-similar expansion~\cite{castin_dum_1996} as shown in Fig.~\ref{fig:tof_b4_after}~(b,c). Similarly the \87Rb BEC is compressed with a flat interface towards the \39K BEC. However, even in the miscible regime, the repulsive interactions are sufficient to affect the spatial distribution of the \39K BEC after expansion as seen in Fig.~\ref{fig:tof_b4_after}~(e,f).

Figure~\ref{fig:dblint_tof_b4_after} shows the corresponding doubly-integrated densities. In the immiscible case (top row) the spatial features arising due to the coupled expansion are clearly visible, both in simulations and experiments. On the one hand, the density of \87Rb exhibits a steep shoulder at the interface with the \39K BEC. Moreover, distinct secondary peaks can be seen in the densities of both components in the directions facing away from the mixture interface. These are caused by the strong repulsion during the initial expansion, which accelerates the two components away from each other. For \39K this effect can be rather pronounced, with the furthest fraction appearing to be almost spatially separated from the main \39K BEC (see also Fig.~\ref{fig:tof_b4_after}~(c)). These features are reminiscent of dispersive shock waves, predicted to arise from the strong mutual expulsion of the two components~\cite{Chang2008, ivanov2017}.

To assess the effect of the thermal clouds we compare our full simulation with a simulation based on the GPE at $T=0$. Figure~\ref{fig:dblint_tof_b4_after} (dashed lines) shows that the presence of the thermal clouds partially suppresses the sharp features in the density distributions after coupled expansion. Nonetheless the same striking features are clearly visible in both simulated and the experimental density distributions.

%*******************************************************************************************************
\subsection{Effects on the centre of mass positions}
\label{sec:COMshift}
%*******************************************************************************************************

The COM position along the axial direction ($z$-axis) is a particularly suitable quantity to investigate the miscible-immiscible transition.

\begin{figure}[htb]
\begin{center}
\includegraphics[width=0.9\textwidth]{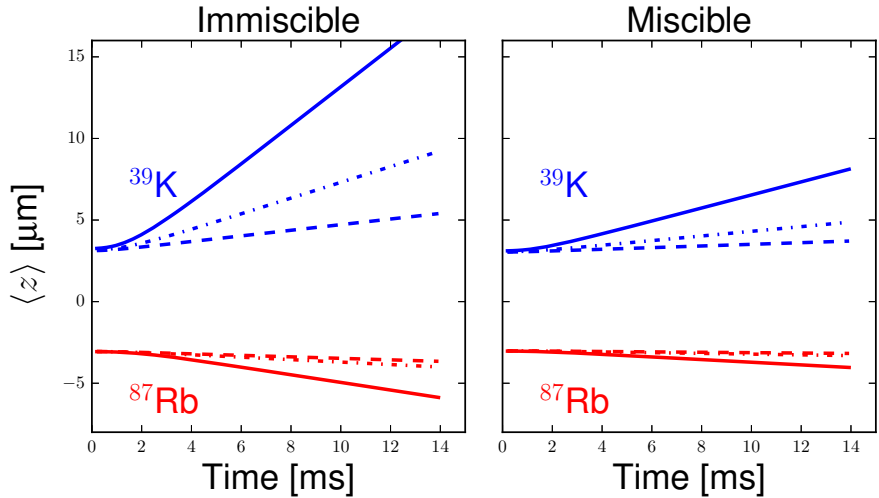}
\caption{Centre of mass position of BECs versus TOF for a mixture of $^{87}$Rb (solid red line) and $^{39}$K (solid blue line) atoms in the immiscible $\Delta = -0.93$ (left) and miscible $\Delta = 1.2$ regimes (right) given in Fig.~\ref{fig:tof_b4_after}. In addition the centre of mass position of the thermal clouds (dashed lines) and the entire clouds (dot dashed lines) are shown. The coordinate systems corresponds to a freely falling frame initially centred between the two trap centres.}
\label{fig:sep_time}
\end{center}
\end{figure}

Following the numerical method outlined above we extract the relative COM positions of binary BECs and the associated thermal clouds after a full simulation of the TOF expansion. Since our computational grid coincides with a moving frame free-falling under gravity our results are given in terms of a relative COM position $\langle z\rangle$. Figure~\ref{fig:sep_time} shows the time evolution of the relative COM positions for all components in the immiscible and the miscible case. In general, the two species accelerate away from each other during the first 6\,ms due to the interspecies repulsion, but travel at an approximately constant speed afterwards as the densities become negligible. We have checked that our simulated COM positions (including other sets of numerical data not shown here) obey the conservation of total momentum.

It is interesting to note that the BEC atoms acquire a greater COM displacement from the interspecies repulsion than the thermal atoms. Since the BEC densities are larger and their distributions are spatially more compact, they experience a larger acceleration due to the interspecies repulsion than the thermal clouds.  

\begin{figure}[htb]
\begin{center}
\includegraphics[width=0.7\textwidth]{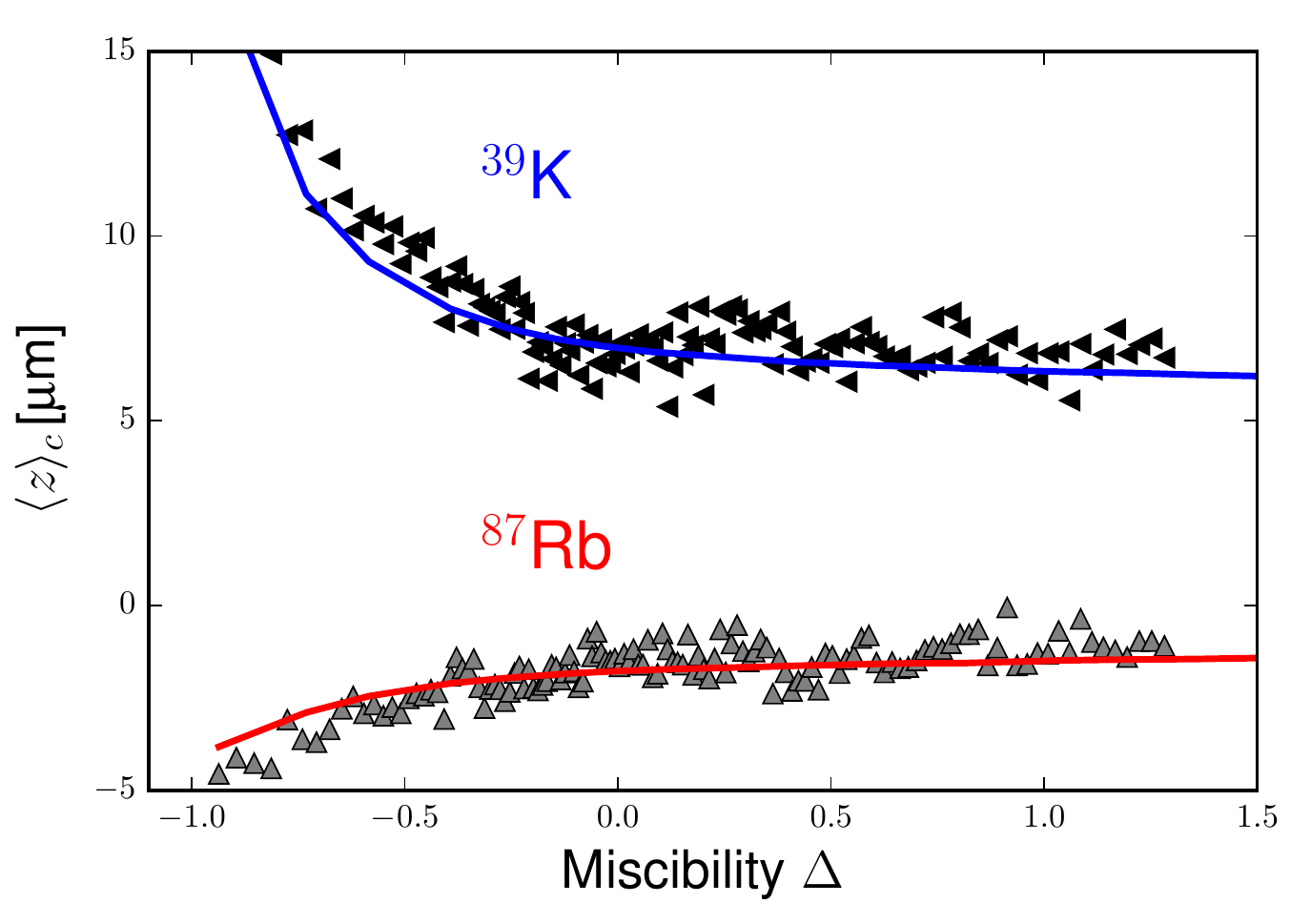}
\caption{Centre of mass position $\langle z\rangle_{\rm c}$ of BECs of $1.2\times10^5$ $^{87}$Rb atoms and $4.2\times10^4$ $^{39}$K atoms after 17\,ms ($^{87}$Rb) and 15\,ms ($^{39}$K) TOF as a function of the miscibility parameter. Experimental realization ($\blacktriangleleft$ and $\blacktriangle$) and simulation with 5.5\,\textmu$ \rm{m}$ offset (solid lines). The origin of the vertical axis corresponds to the position of a freely expanding BEC without the presence the other species.}
\label{fig:sep_sag}
\end{center}
\end{figure}

Figure~\ref{fig:sep_sag} shows the extracted relative COM positions of the BECs compared to experimental data from~\cite{wacker_jorgensen_2015}. In the simulation we extrapolate the COM positions to the experimental time of flight by performing a linear fit to the COM positions between 6\,ms and 14\,ms. We observe good agreement between simulation and experiment, confirming that the qualitative behaviour during TOF is captured well. However, a relatively small offset of 5.5\,\textmu$ \rm{m}$ is required in the simulations in order to reproduce the experimental results. In \ref{sec:role_of_N} we additionally show the qualitative behaviour of the COM positions is independent of the offset.

We have also investigated whether it is physically meaningful to extract a transition point from this relative position. Specifically, we consider the derivative of the relative COM positions with respect to the miscibility, as discussed in more detail in~\ref{sec:role_of_N}. We find no sharp increase of this derivative, which increases monotonically from the miscible to the immiscible regime.

%*******************************************************************************************************
\subsection{Relevance of homogeneous miscibility criterion for trapped mixtures}
\label{sec:insitu_criterion}
%*******************************************************************************************************

To relate our findings to the homogeneous BEC phase-separation criterion, we restrict our discussion here to the case of two pure BECs ($T=0$). Firstly, we recall that the criterion $\Delta=0$ has been derived for a homogeneous $T=0$ \emph{in-situ} mixture at equilibrium~\cite{colson_fetter_1978}, whereas the experimental data corresponds to non-equilibrium expanding density distributions of two BECs which were initially confined in a harmonic trap with different trapping frequencies for the two species. Upon release from the harmonic trap, the BECs expand, with each BEC affected by the mean field of the other one~\cite{hall_matthews_1998}. The shift in relative COM position arises primarily from the interaction within the interface region between the two species. We therefore argue that the transition region is predicted by the mechanical equilibrium of the BEC atoms in this interface region, i.e.
\begin{eqnarray}
\frac{\partial U_{c}^{\rm Rb}}{\partial z} &= g_{\rm Rb-Rb}\frac{\partial n_{c,{\rm Rb}}}{\partial z} + g_{\rm Rb-K} \frac{\partial n_{c,{\rm K}}}{\partial z} &= 0,\nonumber\\
\frac{\partial U_{c}^{\rm K}}{\partial z} &= g_{\rm Rb-K}\frac{\partial n_{c,{\rm Rb}}}{\partial z} \;\,+ g_{\rm K-K} \frac{\partial n_{c,{\rm K}}}{\partial z} &= 0 \;,
\end{eqnarray}
which leads to the usual criterion $\Delta=(g_{\rm Rb-Rb} g_{\rm K-K}/g_{\rm Rb-K}^2) -1 = 0$. This confirms that the agreement of the experimental observations~\cite{hall_matthews_1998, papp_pino_2008, wacker_jorgensen_2015, wang_li_2016} with the homogeneous criterion $\Delta=0$ has its origin in the expansion dynamics, rather than in the \emph{in-situ} density distribution.

%*******************************************************************************************************
\subsection{Miscibility of thermal clouds and BECs}
\label{sec:thermal_bec}
%*******************************************************************************************************

The two species in a mixture will generally have different critical temperatures, implying that the presence of a BEC in one component does not guarantee a BEC in the other. It is therefore interesting to investigate to which extent the presence of a BEC in one component affects the expansion dynamics of a BEC or thermal cloud in the other component coupled to it. This can also shed more light on the miscibility of a binary partially condensed mixture.

In our experiments, the typical range of temperatures and atom numbers is such that it enables us to probe all three cases below in the miscible and immiscible limits:\\

\begin{tabular}{llll}
{\rm Both (partly) condensed:} & $T_{\rm Rb}< T_{\rm c, Rb}$ & {\rm and} & $T_{\rm K}< T_{\rm c, K}$ \\
{\rm Rb (partly) condensed; K thermal:} & $T_{\rm Rb}< T_{\rm c, Rb}$ & {\rm and} & $T_{\rm K} > T_{\rm c, K}$ \nonumber \\
{\rm Both thermal:} & $T_{\rm Rb} > T_{\rm c, Rb}$ & {\rm and} & $T_{\rm K} > T_{\rm c, K}$ \\ \nonumber
\end{tabular}

\begin{figure}[htb]
\begin{center}
\includegraphics[width=\textwidth]{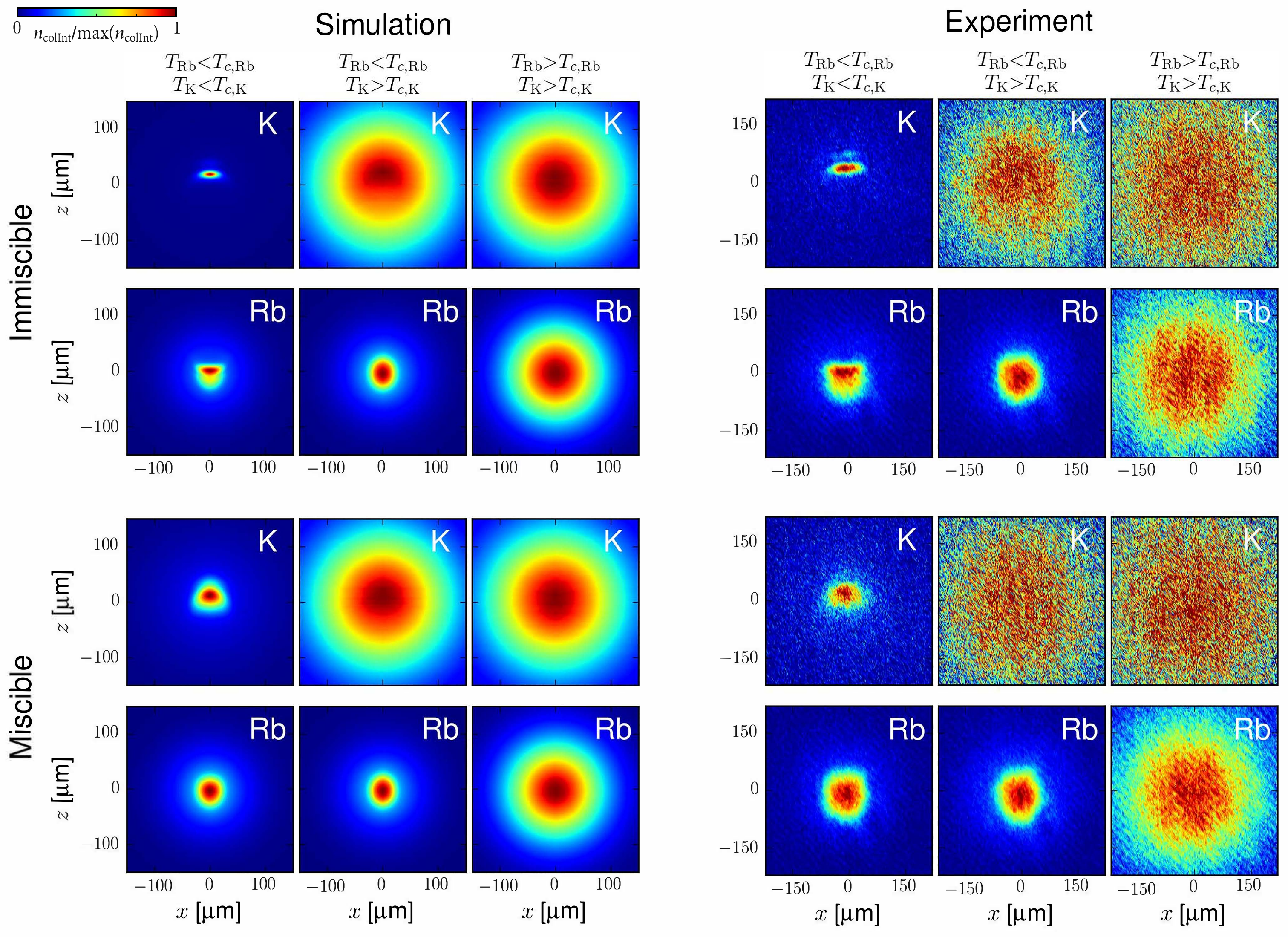}
\caption{Column-integrated density distributions $n_{\rm colInt}$ of $^{39}$K and $^{87}$Rb atoms at different temperatures after 14\,ms of simulated TOF expansion (left columns) compared to corresponding experimental density distributions (right columns) after $\unit[23]{ms}$ ($^{39}$K) and $\unit[24.7]{ms}$ ($^{87}$Rb) TOF expansion. For both simulation and experiment three cases are shown (see text): both species condensed, only $^{87}$Rb is condensed, neither species is condensed. The top two rows represent the immiscible regime, while bottom two rows represent the miscible regime.
In the simulations, the specific parameters considered are as follows (from left to right): 
($T_{\rm Rb} =T_{\rm K} =200\,{\rm nK}$); 
($T_{\rm Rb} =200\,{\rm nK}$ and $T_{\rm K} =270\,{\rm nK}$); and
($T_{\rm Rb} =T_{\rm K} = 270\,{\rm nK}$)
where $T_{c,{\rm Rb}}=222\,{\rm nK}$ and $T_{c,{\rm K}}=231\,{\rm nK}$.
Other parameters and coordinates system as in Fig.~\ref{fig:tof_b4_after}. }
\label{fig:tof_misc_temp}
\end{center}
\end{figure}

The density distributions of the two species after expansion for different miscibilities and temperatures are shown in Fig.~\ref{fig:tof_misc_temp}. A comparison of our numerical simulations (left columns) with the corresponding experimental measurements (right columns) shows excellent agreement for both immiscible (top rows) and miscible (bottom rows) mixtures.

Comparing the three numerical density distributions in the immiscible regime Fig.~\ref{fig:tof_misc_temp}~(top left panels), there is a considerable difference in the axial width and shape of the two expanded systems for different temperature regimes. In particular, if the temperature of \39K is above the critical temperature, the flat repulsive interface disappears in the \87Rb distribution but remains visible in the central part of the thermal \39K distribution. If the temperature of \87Rb is also above the critical temperature, the density distributions are circular in shape, reflecting the isotropic nature of the velocity distribution of the thermal atoms. These features are also present in our experimental TOF images (top right panels).

We also investigate the interplay of miscibility and temperature, by comparing expanded distributions in different temperature combinations for systems expected to be immiscible (Fig.~\ref{fig:tof_misc_temp} top panels) and miscible (bottom panels). When both species are partially condensed, strong features between miscible and immiscible mixtures are observed as discussed in Fig.~\ref{fig:tof_b4_after}. Moreover, we find a slightly-dented $^{39}$K thermal density for the immiscible mixture when only the $^{87}$Rb atoms are partially condensed and no difference when both systems are thermal. This shows that the TOF expansion has little influence on the low density thermal atoms. 

In both immiscible and miscible cases, experimental observations (Fig~\ref{fig:tof_misc_temp} right columns), reveal broadly the same features as in our simulations discussed above. This includes a dented \39K thermal distribution in the presence of a strongly repulsive \87Rb BEC. 

We explain these observations by the difference in the mean-field forces, which are proportional to the density gradients ($\nabla U_c^j$ and $\nabla U_{\rm n}^j$). For example, assuming that the BEC density can be approximated by a Thomas-Fermi distribution $\bigl(\propto (z-z_{{\rm s},k})^2\bigr)$ and the thermal density is approximated by a Gaussian distribution $\bigl(\propto \exp[-m_k\omega_{z,k}^2(z-z_{{\rm s},k})^2/(2k_B T_k)]\bigr)$, we expect the contribution to the mean-field forces to scale like $z-z_{{\rm s},k}$ from a BEC, and $(z-z_{{\rm s},k})\exp[-m_k\omega_{z,k}^2(z-z_{s,k})^2/(2k_B T_k)]$ from a thermal cloud, where the indices $j$ and $k$ refers to the two species. Since the two species will meet and interact at a distance $z$ far from $z_{{\rm s},k}$, we expect the contributions from the BEC to be large while those from the thermal cloud will be minimal due to the extra exponential factor. Hence the presence of the BECs can be inferred from the density distribution of the other species after TOF.

Our analysis confirms that the repelling features seen in Fig.~\ref{fig:tof_b4_after} are a result of the mutual BEC mean field repulsion during expansion (since the original BECs were spatially separated), and that the vertical width of the \39K BEC is significantly compressed by the repulsion. In case only one species is condensed, the difference between miscible and immiscible interactions is less distinguishable.

Although such behaviour is broadly expected, the excellent verification of our numerical model against experiments  enables us to make further numerical predictions about the importance of the trap offset (e.g. due to gravitational sag) in experiments.

%*******************************************************************************************************
\subsection{Effect of the gravitational sag}
\label{sec:sag}
%*******************************************************************************************************

The comparison of our experimental data with the full simulation in the preceding sections has shown that it is vital to appropriately include the initial gravitational sag between the two species. The role of trap sag has previously been theoretically investigated in Refs.~\cite{riboli_modugno_02,jezek_capuzzi_02}, while the effect of temperature has been considered in Refs.~\cite{ohberg_1999,edmonds_lee_2015a,edmonds_lee_2015b,liu_pattinson_2016,lee_proukakis_16}.

\begin{figure}
\begin{center}
\includegraphics[width=0.8\textwidth]{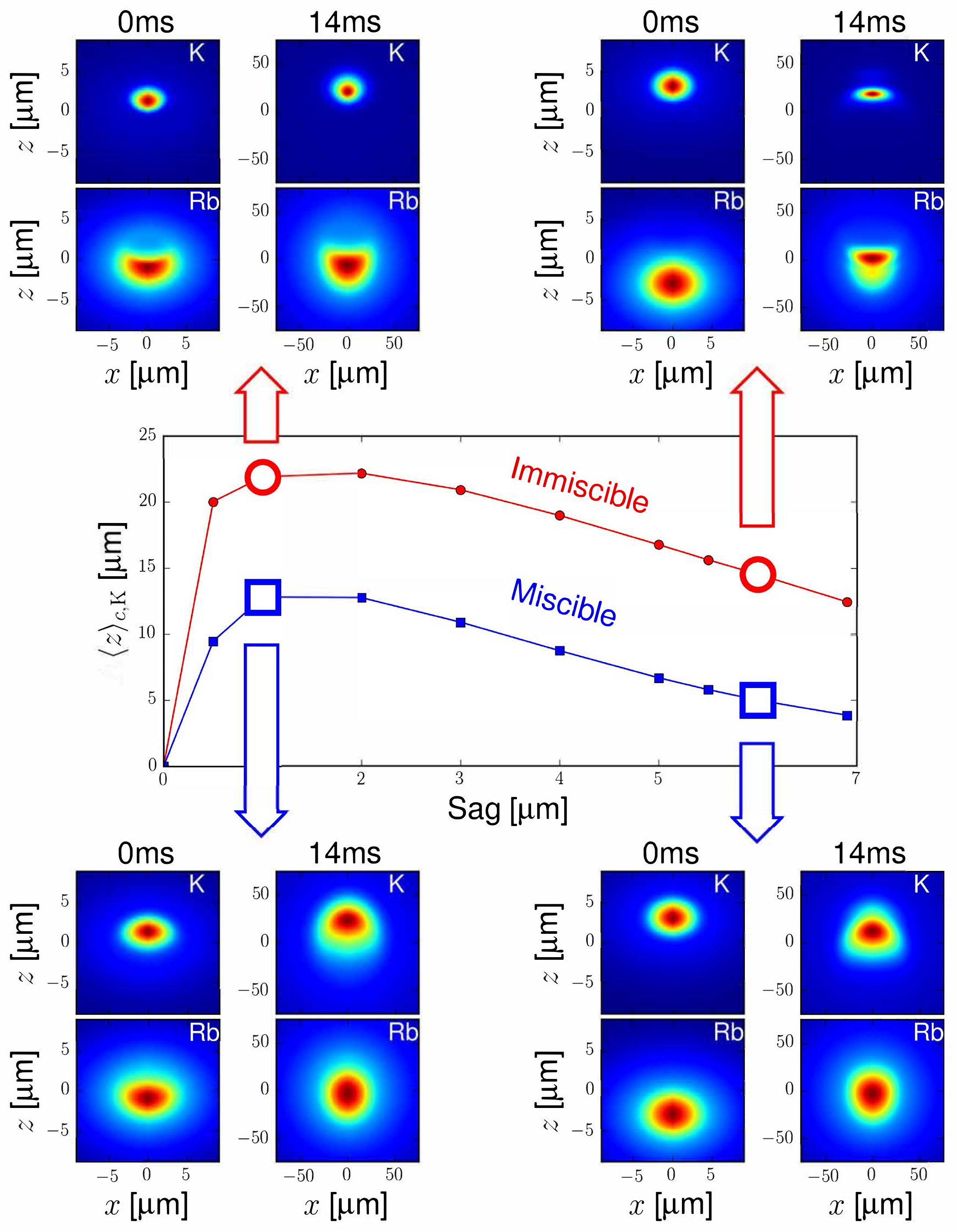}
\caption{Central graph: Centre of mass position  $\langle z\rangle_{\rm c,K}$ of $^{39}$K BECs after expansion for 14~ms as a function of the initial separation between $^{39}$K and $^{87}$Rb BECs. Results of our simulation are shown for both miscible (blue squares) and immiscible (red circles) mixtures, with miscibility parameters given in Fig.~\ref{fig:tof_b4_after}. Corresponding column-integrated density distributions for immiscible (top) and miscible (bottom) cases at separations of $d$ $=1 \, \mu$m (left 2 columns) an $6 \, \mu$m (right 2 columns) for both {\em in situ} and expanded images for both $^{39}$K and $^{87}$Rb BECs. The origin of the vertical axis corresponds to the position of a freely expanding  $^{39}$K BEC without the presence of a  $^{87}$Rb BEC.
}
\label{fig:com_vs_sag_compare_misc}
\end{center}
\end{figure}

Figure~\ref{fig:com_vs_sag_compare_misc} provides a more complete analysis of the effect of the sag on the COM position of the expanded partially condensed $^{39}$K cloud. In both the miscible and the immiscible case, the repulsion of the clouds leads to a shift of the COM position. Not surprisingly this shift is larger for small sag, corresponding to a larger initial overlap of the repulsive clouds, and tends slowly to zero with increasing distance between the two trap centres. This suggests that a small nonzero sag, e.g. one that is considerably smaller than each of the individual Thomas-Fermi radii (see Refs.~\cite{hall_matthews_1998,papp_pino_2008}) is ideal for achieving a maximal relative COM position in experiments. Comparing the COM shift between miscible and immiscible regimes in Fig.~\ref{fig:com_vs_sag_compare_misc} for fixed sag and expansion time, we see a clear enhancement for immiscible clouds, due to the stronger intraspecies scattering length $g_{12}$ which generates a stronger mutually-repulsive potential during expansion.

Our simulations also enable us to address the effect of an initial sag on the density distributions of the expanded clouds. Typical examples of such distributions are shown in Figure~\ref{fig:com_vs_sag_compare_misc} for two different sag values in both immiscible (top) and miscible (bottom) regimes. In each case both the {\em in situ} and the expanded distributions are shown. In the miscible case neither the {\em in situ}, nor the expanded distributions are significantly affected by the sag. Contrary to this, a change in the sag significantly affects the distributions in the immiscible case. Denoting the BEC radii (finite temperature Thomas-Fermi radii) by $R_{\rm Rb}$ and $R_{\rm K}$, and the trap sag by $d$, we distinguish two cases:
(i) For small trap offset $d < R_{\rm Rb} \, , \, R_{\rm K}$ the {\em in situ} distributions already reveal evidence of immiscibility; upon expansion, there is maximum relative COM position, with the curvature of the Rb distribution in the interface region becoming smoother, but not quite planar, while the repelled expanding \39K BEC remains largely spherical.
(ii) Once the trap offset becomes comparable to the sum of the effective BEC radii $d \lessapprox (R_{\rm Rb} + R_{\rm K}$), we see a very different picture: the {\em in situ} distributions are largely decoupled due to their large {\em geometrical} separation, having elliptical shapes (ellipticity set by the trap aspect ratio). However, upon expansion the interface between the two BECs becomes largely planar, and the widths of {\em both} BECs in the vertical direction are highly compressed by the mean-field potentials during expansion. It is precisely those distributions that were analysed in Fig.~\ref{fig:tof_b4_after}, which confirmed the experimental validity of our findings.

We thus conclude that although a small trap sag may be ideal for maximizing the COM displacement between two immiscible BECs, a cleaner signature of the immiscibility of a given mixture is obtained for a larger sag, with the dynamical mean-field pressure upon expansion generating a planar interface between the two initially-disjoint BECs. Moreover the strong acceleration leads to highly non-trivial density distributions with clearly distinguishable side peaks.

%*******************************************************************************************************
\section{Conclusion}
\label{sec:conclusion}
%*******************************************************************************************************

To summarize, we have performed the first {\em ab initio} theoretical analysis of two components which fully includes the mean-field dynamics during TOF expansion, showing very good agreement with experimental results.  Our analysis explains the striking features which emerge in time-of-flight for BECs with strong interspecies repulsion, showing that the expansion can lead to the formation of distinct spatial features, which closely resemble dispersive shock waves~\cite{Chang2008}. Moreover our analysis of the centre of mass positions of the BECs after expansion demonstrates that the homogeneous phase-separation criterion also emerges  in mixtures during expansion as a result of mechanical equilibrium across the BEC interface region. This is an important observation since it reconciles the previously-reported lack of applicability of this criterion for {\em in situ} density distributions of inhomogeneous mixtures with the experimental mixture observations reported to date.  However, we find that no clear transition point between the mixed and the separated phases can be detected in the centre of mass positions after expansion, but that this is a more gradual effect, where $g_{11} g_{22} \approx g_{12}^2 $ marks an approximate transition between miscible and immiscible. The presence of a gravitational sag is found to be advantageous for identifying this transition region. Moreover, we find that this sag should be a fraction of the combined Thomas-Fermi radii of the two BECs to obtain a large shift of the COM position. Finally we analysed the situation where only one component is condensed. In this case distinct features prevail, with the BEC largely symmetric after TOF expansion, but a slightly asymmetric non-condensed species, which is more pronounced in the immiscible case.

This work sheds new light on the evaluation of  multi-component systems after time-of-flight. In particular it shows that the expansion has a major effect on the density distributions and COM position observed after TOF. Thus our analysis will guide future experiments on the detection of miscibility in these systems.

\ack
KLL and NPP acknowledge support from EPSRC Grant No. EP/K03250X/1. This work made use of the facilities of N8 HPC Centre of Excellence, provided and funded by the N8 consortium and EPSRC (Grant No.EP/K000225/1). The Centre is co-ordinated by the Universities of Leeds and Manchester. The Aarhus group acknowledges support by the Lundbeck Foundation, the Villum foundation and the Independent Research Fund Denmark.

Data supporting this publication is openly available under an `Open Data Commons Open Database License'. Additional metadata are available at: http://dx.doi.org/10.17634/ 122626-5. Please contact Newcastle Research Data Service at rdm@ncl.ac.uk for access instructions.

%*******************************************************************************************************
\newpage
\appendix
%*******************************************************************************************************
\section*{Appendix}
%*******************************************************************************************************
\section{Numerical implementation of the ``ZNG'' scheme in cylindrical coordinates}
\label{sec:app_numerical}
%*******************************************************************************************************

%*******************************************************************************************************
\subsection{Equilibrium solution}
%*******************************************************************************************************

Following our earlier work \cite{edmonds_lee_2015a, edmonds_lee_2015b, lee_proukakis_16}, our numerical simulation begins by finding the equilibrium distribution of the mixture at finite temperatures. Because of the gravitational sag, the atomic clouds of the two species can have a small spatial overlap, which can lead to inefficient interspecies thermalisation. We thus allow the possibility that the two species have two distinct temperatures $T_j$, but atoms of the same species are at thermal equilibrium with each other.

The equilibrium calculations proceed by computing the semi-classical Hartree-Fock approximation of the local thermal atom density
\begin{equation}\label{eq:eq_ntherm}
\tilde{n}_j(\vec{r})= g_{3/2}(z_j)/\lambda_j^3
\end{equation}
with thermal wavelength $\lambda_j = \sqrt{2\pi\hbar^2/(m_j k_B T_j)}$, local fugacity $z_j = \exp[(\mu_j-U_n^j )/(k_B T_j )]$ and the textbook result of Bose function,
\begin{equation}
g_{3/2}(z)=\frac{2}{\sqrt{\pi}} \int_0^\infty dx \frac{\sqrt{x}}{\Exp{x}/z-1}.
\end{equation}
The chemical potential $\mu_j$ is obtained from the imaginary-time propagation of \eref{eq:gpe}. We then obtain the equilibrium BEC and thermal density distributions by solving equations \eref{eq:gpe} and \eref{eq:eq_ntherm} self-consistently. 

%*******************************************************************************************************
\subsection{Dynamical expansion}
%*******************************************************************************************************

To model the dynamical evolution, equations \eref{eq:gpe} and \eref{eq:qbe} are typically solved on a position-space computation grid with regular grid spacings. However, it is computationally prohibitive to solve these equations for the expansion dynamics on a three-dimensional computational grid with the same grid spacings throughout the dynamical simulation. A possible solution is to increase the grid spacings with time such that it is possible to use a limited number of grid points to cover the spatial extend of the expanding atomic clouds. This amounts to truncating the high-wavenumber component in the wave functions/densities and studying the densities at a coarser length scale. In our case, this approach is not suitable as the interspecies repulsion leads to non-negligible average velocities for each species. The grid spacings therefore have to be small enough to capture the wavelength associated with these average velocities.

By utilizing the cylindrical symmetry of the problem, we can solve the same equations with only two dimensions, the radial ($\rho=\sqrt{x^2+y^2}$) and the axial ($z$) dimensions. With the same number of grid points as a typical three-dimensional simulations (e.g. $128^3$ or $256^3$), we are able to study the atomic clouds that have expanded up to approximately ten times their in-trap sizes without changing the grid spacings. More details on our numerical approach to solve equations \eref{eq:gpe} and \eref{eq:qbe} within the cylindrical polar coordinate system are given below.

If the atomic clouds are not rotating about the axial direction, the wave functions $\phi_j(\rho,z)$ and the density distributions $\bigl(n_{c,j}(\rho,z)$ and $\tilde{n}_j(\rho,z)\bigr)$ depend only on the radial ($\rho$) and the axial ($z$) coordinates when the cylindrical symmetry is present. 

The GPE~\eref{eq:gpe} is therefore reduced to 
\begin{equation}
i\hbar\frac{\partial\phi_j}{\partial t}=-\frac{\hbar^2}{2m_j}\bigg[\frac{\partial^2}{\partial\rho^2}+\frac{1}{\rho}\frac{\partial}{\partial\rho}+\frac{\partial^2}{\partial z^2}\bigg]\phi_j+U^{j}_{c}\phi_j.\label{eq:cylin_gpe}
\end{equation}
It can be discretized on a two-dimensional spatial grid,
\begin{eqnarray}
		\rho_l\,\, =& \left(l-\frac{1}{2} \right)\times d\rho \quad & \textrm{for}\quad l=1,2,\ldots,N_\rho,\\
		z_m =& \left(-\frac{N_z}{2} + m\right)\times dz \quad & \textrm{for}\quad m = 1,2,\ldots,N_z.
\end{eqnarray}

\noindent with the radial grid spacing $d\rho$, the axial grid spacing $dz$ and a total of $N_\rho\times N_z$ grid points. The time-propagation is then solved with alternating direction implicit method. In particular, we use the Crank-Nicolson method~\cite{crank_nicolson_1947} for the radial direction and the Fourier spectral split-step method~\cite{antoine_bao_2013} for the axial direction.

In order to apply the Crank-Nicolson method, we approximate the radial differential operators with a finite-difference approximation. Using the notation $\phi_j(l,m) \equiv \phi_j(\rho_l,z_m)$, we arrive at
\begin{eqnarray}
		\label{eq:diff_rho}\frac{1}{\rho}\frac{\partial\phi_j}{\partial\rho}\bigg|_{\rho=\rho_l,z=z_m} &\approx& \frac{1}{\rho_j(l,m)}\frac{\phi_j(l+1,m)-\phi_j(l-1,m)}{ 2d\rho}. \\
		\label{eq:laplacian_rho}\frac{\partial^2\phi_j}{\partial\rho^2}\bigg|_{\rho=\rho_l,z=z_m} &\approx & \frac{\phi_j(l+1,m)-2\phi_j(l,m)+\phi_j(l-1,m)}{d\rho^2}.
\end{eqnarray}

\noindent We further choose $\phi_j(0,m)=\phi_j(1,m)$ to enforce the boundary condition $\partial\phi_j/\partial\rho=0$ at $\rho=0$ such as to preserve the cylindrical symmetry. Equations~\eref{eq:diff_rho} and \eref{eq:laplacian_rho} then become exact at $\rho=d\rho/2$ if $\phi_j=c_0 + c_2\rho^2$ for any constant $c_0$ and $c_2$.

On the other hand, the quantum Boltzmann equation~\eref{eq:qbe} is solved with the direct simulation Monte Carlo method~\cite{bird_1976,bird_1994}. A large number of test particles (typically of the order of millions) are sampled with the acceptance-rejection method according to the Bose-Einstein distribution $[\Exp{(\varepsilon^j_{\vec{p}}-\mu_j)/(k_BT_j)}-1]^{-1}$, where $\varepsilon^j_{\vec{p}}=\frac{\vec{p}^2}{2m_j} + U_{\rm n}^j$ is the local energy of a thermal atom. These test particles are then evolved in time according to Newton's equation of motion using the symplectic leapfrog method. 

In contrast to the BEC propagation, simulating the test particle dynamics with cylindrical polar coordinates requires a computational grid with irregular radial grid spacings, where the radial grid points $\tilde{\rho}_l$ are located at a constant multiple of the zeros $\zeta_l$ of the Bessel function $J_0$. The main reason is the need for an efficient numerical method to compute the discrete Hankel transformation (see details in the next paragraph). Arranging the zeros $\zeta_l$ in increasing order, with $\zeta_1$ being the smallest zero ($\zeta_1=2.4048$), we have 
\begin{equation}
\tilde{\rho}_l = \frac{\zeta_{l}}{\zeta_{N_\rho+1}}\rho_{\rm max},\quad{\rm for}\quad l=1,2,\ldots,N_\rho\label{eq:th_rgrid}
\end{equation}
where $\rho_{\rm max}$ gives the radial boundary of the computational grid.

Three technical points are worth mentioning here: (1) binning of the test particles into spatial cells, (2) smoothing of the resulting distributions so as to approximate $\tilde{n}_j$ and (3) computing the mean-field force acting on the test particles from the spatial derivative of the effective potential $U_{\rm n}^j$.

\begin{enumerate}
\item \emph{Binning of test particles}\\
Suppose that we are simulating the dynamics of $\tilde{N}_j$ thermal atoms using $N_{{\rm tp}, j}$ test particles. Each test particle therefore carries a weight $w_j = \tilde{N}_j / N_{{\rm tp},j}$. For a test particle with a position vector $(x,y,z)\equiv (\rho,z)$, we assign its contribution to $D_j(\rho,z) \equiv 2\pi\rho\,\tilde{n}_j(\rho,z)$ using the cloud-in-cell method: for $\tilde{\rho}_{l-1}<\rho<\tilde{\rho}_l$ and $z_{m-1}<z<z_m$, we add the contribution 
\begin{eqnarray}
		D_{j,{\rm approx}}(\tilde{\rho}_{l-1},z_{m-1}) &= &\frac{w_j}{A_{l-1}} \left(\frac{\tilde{\rho}_l-\tilde{\rho}}{\tilde{\rho}_l-\tilde{\rho}_{l-1}}\right)\left(\frac{z_m-z}{z_m-z_{m-1}}\right),\quad\quad\;\; \\
		D_{j,{\rm approx}}(\tilde{\rho}_{l-1},z_{m}) &= & \frac{w_j}{A_{l-1}} \left(\frac{\rho_l-\rho}{\tilde{\rho}_l-\tilde{\rho}_{l-1}}\right)\left(\frac{z-z_{m-1}}{z_m-z_{m-1}}\right),\quad\quad\;\;\\
		D_{j,{\rm approx}}(\tilde{\rho}_{l},z_{m-1}) &= & \frac{w_j}{A_{l}} \left(\frac{\rho-\rho_{l-1}}{\tilde{\rho}_l-\tilde{\rho}_{l-1}}\right)\left(\frac{z_m-z}{z_m-z_{m-1}}\right),\quad\quad\;\;\label{eq:d10}\\
		D_{j,{\rm approx}}(\tilde{\rho}_{l},z_{m}) &= & \frac{w_j}{A_{l}} \left(\frac{\rho-\rho_{l-1}}{\tilde{\rho}_l-\tilde{\rho}_{l-1}}\right)\left(\frac{z-z_{m-1}}{z_m-z_{m-1}}\right), \label{eq:d11} 
\end{eqnarray}

\noindent where $A_l=(\tilde{\rho}_{l+1}-\tilde{\rho}_{l-1})/2\times dz$ is the area element associated with the radial grid point $\rho_l$. In the case that $\rho<\tilde{\rho}_1$, we only compute equations \eref{eq:d10} and \eref{eq:d11}. The thermal atom density is then approximated by $\tilde{n}_{j,{\rm approx}}(\rho,z)=D_{j,{\rm approx}}(\rho,z)/(2\pi\rho)$.
\item \emph{Smoothing of approximated $\tilde{n}_j$}\\
The approximated $\tilde{n}_{j,{\rm approx}}$ contains undesirable and spurious spatial fluctuations due to the binning procedure (see e.g. Fig.~1 of~\cite{jackson_zaremba_2002}). These fluctuations can be reduced by convolving $\tilde{n}_{j,{\rm approx}}$ with a Gaussian filter $G(\vec{r})=(\pi\sigma^2)^{-3/2}\Exp{-\vec{r}^2/\sigma^2}$. In its three-dimensional form, we compute
\begin{equation}
\tilde{n}_{j,{\rm conv}}(\vec{r}) = \int d\vec{r}'\; G(\vec{r}-\vec{r}')\;\tilde{n}_{j,{\rm approx}}(\vec{r}').
\end{equation}
This can be done efficiently by employing Fourier transforms and the convolution theorem. For our cylindrically-symmetric problem, the numerical Fourier-transformation amounts to performing a discrete Fourier transforms for the axial direction and a discrete Hankel transformations for the radial direction. While the former can be conveniently performed using, e.g. the FFTW library~\cite{frigo_1999}, we employ the quasi-discrete Hankel transforms~\cite{guizar_sicairos_gutierrez_vega_2004} for the latter. Remarkably, the convolution theorem remains applicable under the discrete Hankel transformation~\cite{baddour_2009}.

To be concrete, a two-dimensional Fourier transform of a radial function $R(\rho)$ is reduced to a zeroth-order Hankel transform,
\begin{eqnarray}
		K(\kappa)&=&\int_0^\infty\, d\rho \, \rho \int_0^{2\pi} d\theta\, \Exp{\mri \kappa\rho\cos\theta}R(\rho) \nonumber\\
             &=&2\pi\int_0^\infty\, d\rho \, \rho\, R(\rho)J_0(\kappa\rho)
\end{eqnarray}
The corresponding inverse transform is
\begin{eqnarray}
	R(\rho)&=&\frac{1}{(2\pi)^2}\int_0^\infty\, d\kappa \, \kappa \int_0^{2\pi} d\theta\, \Exp{-\mri \kappa\rho\cos\theta}K(\kappa) \nonumber\\
         &=&\frac{1}{2\pi}\int_0^\infty\, d\kappa \, \kappa\, K(\kappa)J_0(\kappa\rho)
\end{eqnarray}
If we choose to evaluate $R(\rho)$ at $\rho=\tilde{\rho}_l$~\eref{eq:th_rgrid} and $K(\kappa)$ at $$\kappa=\kappa_m=\frac{\zeta_{m}}{2\pi \rho_{\rm max}}$$ for $m=1,2,\ldots,N_{\rho}$, the two vectors, $R(\tilde{\rho}_l)$ and $K(\kappa_m)$, are related through
\begin{eqnarray}
\mathcal{R}(\tilde{\rho}_l) &=& \sum_{m} \mathcal{T}_{lm}\mathcal{K}(\kappa_m), \nonumber\\
\mathcal{K}(\kappa_m) &=& \sum_{l} \mathcal{T}_{ml}\mathcal{R}(\tilde{\rho}_{l}),
\end{eqnarray}
where
\begin{eqnarray}
\mathcal{R}(\tilde{\rho}_l) &=& R(\tilde{\rho}_l) \frac{\rho_{\rm max}}{J_1(\zeta_{l})},\nonumber\\
\mathcal{K}(\kappa_m) &=& K(\kappa_m) \frac{\zeta_{\rm N_\rho+1}}{2\pi\rho_{\rm max} J_1(\zeta_{m})},\nonumber\\
\mathcal{T}_{lm} &=& \frac{2 J_0(\zeta_l \zeta_m / \zeta_{N_\rho+1})}{|J_1(\zeta_l)|J_1(\zeta_m)|\zeta_{N_\rho+1}}.
\end{eqnarray}
In particular, $\mathcal{T}$ is a real square symmetric matrix, which becomes nearly unitary when $N_\rho\gg1$. In our simulation, $N_{\rho}=1000$.

\item \emph{Computing the mean-field force}\\
After we have smoothed the binned density, we map $\tilde{n}_{j,{\rm conv}}$ from the irregular radial grid $\tilde{\rho}_l$ to the regular radial grid $\rho_l$ used in BEC propagation through linear interpolation. The interpolated density is then used to compute the effective potential $U_{{\rm n}}^j$. \\

In order to solve Newton's equation of motion, we estimate the force 
\begin{equation}
	{\bf F}=-\nabla U_{\rm n}^j = -\frac{\partial U_{\rm n}^j}{\partial \rho}\hat{\bm{\rho}} - \frac{\partial U_{\rm n}^j}{\partial z}\hat{\bm{z}}
\end{equation}
through a finite difference scheme and linear interpolation. For a test particle with a position vector $(x,y,z)\equiv(\rho,z)$, where $\rho_{l-1}<\rho<\rho_l$ and $z_{m-1}<z<z_m$, we have
\begin{eqnarray}
		\frac{\partial U_{\rm n}^j}{\partial \rho} &\approx& \quad\frac{U_{\rm n}^j(\rho_{l},z_{m-1})-U_{\rm n}^j(\rho_{l-1},z_{m-1})}{d\rho} \frac{z_m-z}{dz}\nonumber\\
                                               &&+ \frac{U_{\rm n}^j(\rho_{l},z_m)-U_{\rm n}^j(\rho_{l-1},z_m)}{d\rho}\frac{z-z_{m-1}}{dz},\\
\frac{\partial U_{\rm n}^j}{\partial z} &\approx& \quad\frac{U_{\rm n}^j(\rho_{l-1},z_{m})-U_{\rm n}^j(\rho_{l-1},z_{m-1})}{dz} \frac{\rho_{l}-\rho}{d\rho}\nonumber\\
                                               &&+ \frac{U_{\rm n}^j(\rho_{l},z_m)-U_{\rm n}^j(\rho_{l},z_{m-1})}{dz}\frac{\rho-\rho_{l-1}}{d\rho}.                                                              
\end{eqnarray}
\end{enumerate}

%*******************************************************************************************************
\section{Changing or not changing the interaction strength after 6\,ms of TOF expansion}
\label{sec:role_of_quench} 
%*******************************************************************************************************

In our experiments the magnetic field is swept to a low value after 6~ms of TOF expansion for imaging purpose. This means that the scattering lengths experienced by the system are changed to $a_{\rm Rb-K}=28a_0$ and $a_{\rm K-K}=-50a_0$. We have simulated the situation including this change of scattering length as well as the situation where the scattering lengths remain unchanged for 14\,ms of expansion time. We find only negligible differences in the resulting COM positions. Our simulation does not include the fact that two Feshbach resonances are crossed as the magnetic field is lowered. However, the good overall agreement between theory and experiment indicates that, these effects are negligible for the low densities of the samples after 6~ms TOF expansion. For all simulation results presented in the main text the change of scattering length after 6~ms of TOF expansion was included.

%*******************************************************************************************************
\section{Characterization of the miscible--immiscible crossover}
\label{sec:role_of_N}
%*******************************************************************************************************

\begin{figure}[h]
\begin{center}
\includegraphics[width=0.65\textwidth]{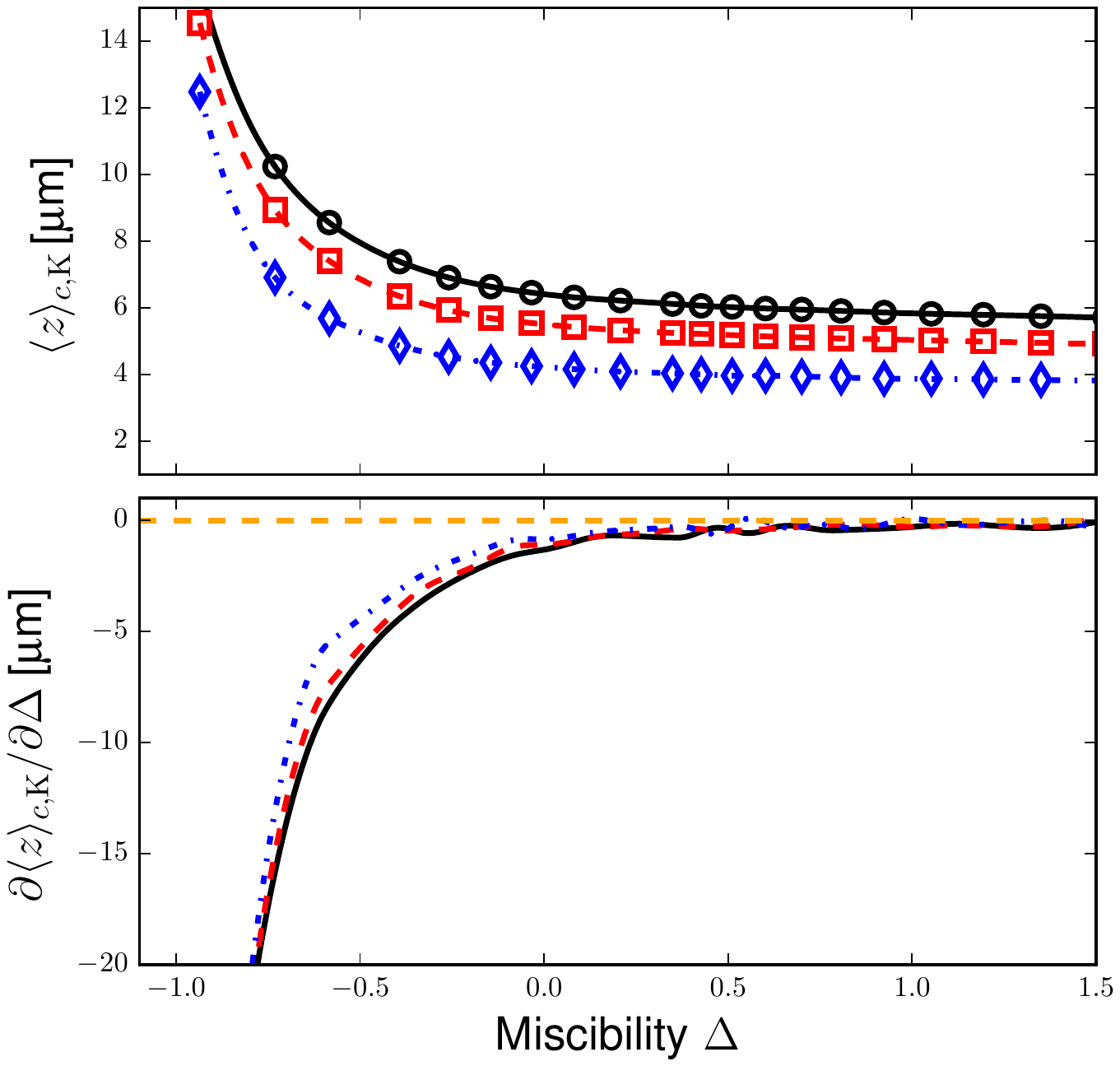}
\caption{ Top: Centre of mass position of the $^{39}$K BEC as a function of the miscibility $\Delta$ for different trap offsets, after 14\,ms of TOF expansion with 6.9$\mu$m offset ($\lozenge$), 6$\mu$m offset ($\square$), 5.5$\mu$m offset ($\circ$). The lines correspond to cubic-spline interpolation. Bottom: Derivative of the position with respect to the miscibility, based on the cubic spline interpolation. The orange dashes indicates zero. To show the observable shift of the COM position, the initial trap offset was subtracted form the theoretical results.}
\label{fig:com_shift_derivative_Rb1.2e5K4.2e4}
\end{center}
\end{figure}

To investigate whether it is physically meaningful to extract a transition point from the relative COM position of expanded clouds, we interpolate our simulated data points by cubic splines (Fig.~\ref{fig:com_shift_derivative_Rb1.2e5K4.2e4} top) and extract the first-order derivative of the relative COM positions with respect to the miscibility (Fig.~\ref{fig:com_shift_derivative_Rb1.2e5K4.2e4} bottom). A well defined transition point should manifest itself as a sharp increase of the derivative. While the relative shift is continuously increasing from the miscible ($\Delta>0$) to the immiscible ($\Delta<0$) regime, the derivative does not show a sharp transition point. 

Fig.~\ref{fig:com_shift_derivative_Rb1.2e5K4.2e4} also shows that the expansion dynamics depends on the chosen offset. The best agreement between theory and experiment is obtained for an offset of 5.5\,\textmu$ \rm{m}$ in our simulation, despite the fact that the estimated experimental offset is 6.9\,\textmu$ \rm{m}$. Several sources contribute to this discrepancy, including inaccurate knowledge of the precise scattering lengths, a systematic error in the evaluation of the centre of mass condensate positions and incomplete knowledge about the precise trap offset.

\bibliography{binaryTOF}

\end{document}